\documentclass[preprint,journal]{vgtc}       

\ifpdf
  \pdfoutput=1\relax                   
  \usepackage{graphicx}                
  \DeclareGraphicsExtensions{.pdf,.png,.jpg,.jpeg} 
\else
  \usepackage{graphicx}                
  \DeclareGraphicsExtensions{.eps}     
\fi%

\usepackage{microtype}                 
\PassOptionsToPackage{warn}{textcomp}  
\usepackage{textcomp}                  
\usepackage{mathptmx}                  
\usepackage{times}                     
\usepackage{cite}                      
\usepackage{tabu}                      
\usepackage{booktabs}                  

\usepackage{balance}
\usepackage{amsmath}
\usepackage{amssymb}
\usepackage{xcolor}
\usepackage{gensymb}

\usepackage{tikz}
\usepackage{pgfplots}
\pgfplotsset{compat=1.12}
\usepgfplotslibrary{fillbetween}

\usepackage[ruled,boxed,vlined,linesnumbered]{algorithm2e}

\newcommand{\etal}{\emph{et al.}}

\graphicspath{{figures/}{figures_lowres/}{pictures/}}

\onlineid{1137}

\vgtccategory{Research}
\vgtcpapertype{algorithm/technique}

\title{Interactive Visualization of Atmospheric Effects for Celestial Bodies}

\author{Jonathas Costa, Alexander Bock, Carter Emmart, \\Charles Hansen, \textit{Fellow, IEEE}, Anders Ynnerman and Cl\'audio Silva, \textit{Fellow, IEEE}}
\authorfooter{
  \item Jonathas Costa and Cl\'audio Silva are with New York University. E-mail: \{jccosta, csilva\}@nyu.edu
  \item Alexander Bock and Anders Ynnerman are  with Link\"oping University and the University of Utah. E-mail: \{alexander.bock, anders.ynnerman\}@liu.se
  \item Carter Emmart is with the American Museum of Natural History. E-mail: carter@amnh.org.
  \item Charles Hansen is with the University of Utah. E-mail: hansen@cs.utah.edu.
  }

\shortauthortitle{Costa \etal: Interactive Visualization of Atmospheric Effects for Celestial Bodies}

\abstract{
We present an atmospheric model tailored for the interactive visualization of planetary surfaces. As the exploration of the solar system is progressing with increasingly accurate missions and instruments, the faithful visualization of planetary environments is gaining increasing interest in space research, mission planning, and science communication and education. Atmospheric effects are crucial in data analysis and to provide contextual information for planetary data. Our model correctly accounts for the non-linear path of the light inside the atmosphere (in Earth's case), the light absorption effects by molecules and dust particles, such as the ozone layer and the Martian dust, and a wavelength-dependent phase function for Mie scattering. The mode focuses on interactivity, versatility, and customization, and a comprehensive set of interactive controls make it possible to adapt its appearance dynamically. We demonstrate our results using Earth and Mars as examples. However, it can be readily adapted for the exploration of other atmospheres found on, for example, of exoplanets. For Earth's atmosphere, we visually compare our results with pictures taken from the International Space Station and against the CIE clear sky model. The Martian atmosphere is reproduced based on available scientific data, feedback from domain experts, and is compared to images taken by the Curiosity rover. The work presented here has been implemented in the OpenSpace system, which enables interactive parameter setting and real-time feedback visualization targeting presentations in a wide range of environments, from immersive dome theaters to virtual reality headsets.
}


\CCScatlist{ 
 \CCScat{K.6.1}{Management of Computing and Information Systems}%
{Project and People Management}{Life Cycle};
 \CCScat{K.7.m}{The Computing Profession}{Miscellaneous}{Ethics}
}

\teaser{
  \centering
  \includegraphics[width=\linewidth]{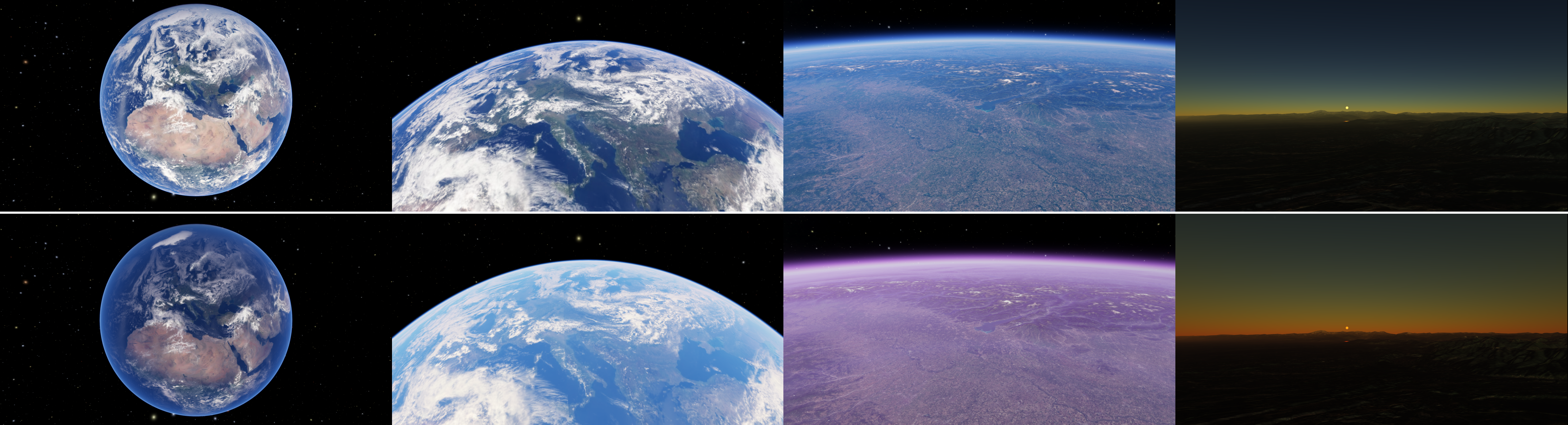}
  \caption{Faithful representation of Earth's atmosphere with our atmospheric model. The atmospheric composition can be changed to simulate exoplanetary atmospheres or, in this case, be based on measured data to represent atmospheres accurately. From left to right, comparing the bottom row with the top row, we can see Earth's atmosphere with 6x more dust particles, 20x higher polarizability, 0.5x higher air's refractive index for wavelengths of 680 nm, and 2x denser.}
	\label{fig:teaser}
}

\vgtcinsertpkg

\begin{document}


\firstsection{Introduction}
\maketitle


The fundamental role of visualization in the exploration of scientific data has over the years become indisputable, and documented successes of visualization as an enabling technology across many disciplines create a solid foundation for the research field. Interactive data visualization also serves an increasingly important role in science communication and general outreach to broad audiences~\cite{ynnerman:16:inside}. The use of exploratory visualization in public spaces such as museums and science centers is elaborated upon by Ynnerman~\etal ~\cite{ynnerman:exploranation}. Recently, advances in computer hardware, increased availability of data, and improvements in visualization methodology have changed the traditional approach to science communication based on explanatory visualization to include elements of interactive data exploration.
The neologism ``Exploranation'' denotes this convergence of exploration and explanation~\cite{ynnerman:exploranation}. Additionally, the need for explanatory approaches in scientific data exploration and documentation is also increasing as interactive visualization is maturing in many application domains as part of the workflow. The software OpenSpace~\cite{OpenSpace:2020} for astrovisualization is positioned in the middle of this ongoing evolution of science communication. It can be used as an interactive science communication tool for live presentations in situations ranging from planetarium and dome theater shows to classroom teaching and individual exploration on personal computers. It is, however, also used as an exploration tool in space and astronomy research in conjunction with other analysis tools and indeed has shown to be powerful in team communication and collaboration. 

An important aspect of visualization at this intersection of exploration and explanation is the need for accurate visual context. In traditional science communication, explanatory visualization context and use of realism in visual metaphors provide real-world context. It has been shown that for novice users and learners context and completeness significantly aids in the formulation of mental models~\cite{KOZMA2003205}. This need carries over to the use of interactive data exploration for broad audiences and calls for significant emphasis on context and realism. The notion of scientific exploration is based on data-driven approaches, while at the same time, contextual content that maintains realism must also be based on data visualization. Inevitably this means that there is a need for approaches to visual representations that are positioned at the intersection of realistic graphics and data visualization. In this paper we present such an approach and present how realistic looking atmospheres can provide context and support interactive exploration of data from space exploration missions and astronomical observations. As the developed model and visualization is data-driven and configurable, it itself is a tool for exploration of the parameter spaces involved in light transport in differing planetary atmospheres and an example of how the ``exploranation'' process works in both ways.  
To achieve this, we implemented a highly configurable parametric system inside OpenSpace~\cite{OpenSpace:2020}. The domain expert can interactively calibrate the physical parameters of the atmosphere based on sensed or computed atmospheric data and visualize the results in real-time from any angle and position. 

Our paper has three main contributions. First, we propose a novel atmospheric model whose main goal is to support a unified parameterized model capable of rendering the atmospheres of different planets and exoplanets with high accuracy. Second, we give an efficient open-source implementation of our new model that is able to run at interactive rates on existing graphics hardware. Finally, we validate our model with real-world data (when available) from Earth's and Mars' atmospheres as well as real-world images from these planets. The validation is based on luminance results from Earth's atmosphere for different Sun's positions are plotted against the theoretical model provided by the International Commission on Illumination (CIE)~\cite{Darula:2002}. 


\section{Related Work}


The faithful visualization of planetary atmospheres has been a long time focus of the scientific community. The atmospheric effects can improve the understanding of the Earth and other planetary systems and help to understand different techniques of realistic visualization, providing the appropriate context for them~\cite{vis17-bladin-globe-browsing}. In this scenario, realistic atmospheric effects are the main aim for scientists working with real-time astronomic visualization and education aiding.

\subsection{Atmospheric Models and Rendering}

Modeling realistic rendering of atmospheres is a vast field and can be classified by two orthogonal types: offline/real-time and single/multiple scattering.


\noindent \textbf{Offline Single Scattering} \quad Nishita~\etal~\cite{Nishita:1993} were one of the first to approach the problem of rendering realistic atmospheres. In their work, the path of a light ray through the atmosphere and its energy contribution is calculated by numerical integration on the CPU that is partially cached in lookup tables. Their model does not account for multiple scattering, prohibit height-dependent atmospheric density changes, and the atmosphere can only be rendered from the outside.

The work of Riley~\etal~\cite{Riley:2004} proposes an analytical solution through the use of multiple scattering phase functions to model the angular dependency of the scattering events (Mie scattering effects are simulated by the use of specific phase functions for water vapor and dust). In Riley's solution, only single scattering effects were considered for an observer inside the atmosphere (on the ground). Also, no transitions between the ground and space were considered.

\noindent \textbf{Real-Time Single Scattering} \quad Hoffman~\etal's~\cite{Hoffman:2002} work extends the solution by Nishita to provide interactive frame rates. They considered the Earth's atmosphere density constant and made strong approximations of the radiative transfer equation~\cite{Chandrasekhar:1960, Mishchenko:2002} (exchanging integral calculations by multiplication of factors).

 Based on the work of Nishita~\cite{Nishita:1993}~\etal, O'Neil~\cite{ONeil2004} proposed an algorithm to generate a real-time approximation of the Earth's atmosphere. He used a 2D lookup table for storing the optical depth and a 3D lookup table to store a precomputed single scattering integral. In O'Neil's next work~\cite{ONeil:2005}, he eliminated the lookup tables. He approximated the single scattering equations by evaluating the integrals using a low sampling approach in the shader, using the graphics hardware to interpolate the atmosphere's final colors value. This latter method obtained good results in real-time and was also capable of rendering the atmosphere from any viewpoint on the ground or in space.

The model by Schafhitzel~\etal~\cite{Schafhitzel:2007} also uses a precomputation model of the single scattering integral, and stores it as a lookup table (as suggested by O'Neil), thus accessing it by a smaller number of parameters (the view and Sun zenith angle). The small number of parameters in the parametrization of the table (the angle between the sun position and the view direction is not available in their solution) prevents the final model from displaying the Earth's shadow inside the atmosphere.


To correctly reproduce colors in the twilight, multiple scattering of the light must be considered as the light traverses more atmosphere during the twilight.

\noindent \textbf{Offline Multiple Scattering} \quad Following their previous work, Nishita~\cite{Nishita:1996} used volume radiosity methods that divide the sky hemisphere in cells to take multiple light scattering effects into account. This enables the generation of images for observers inside or outside the atmosphere, and produces visually suitable images but takes a long time to generate a static image as it consumes a large amount of computational resources and is thus not suitable for a real-time system.

Applying a Monte-Carlo simulation technique on an analytical model, Preetham~\etal~\cite{Preetham:1999} presented a solution for the multiple scattering problem. Their model produces good images for an observer on the ground but is not capable of generating images for an observer outside the atmosphere. Also, as Zotti~\etal~pointed out in~\cite{Zotti:2007}, the model has some incorrect behavior producing negative intensities for some cases.

Horsek~\cite{Hosek:2012}~\etal~proposed an analytical atmospheric model that improves Preetham's model~\cite{Preetham:1999}. They improved the final analytical formula by adding more degrees of freedom for the fitting phase. They added new parameters to the model besides the turbidity (an indirect measure of the number of suspended particles in a medium) and executed all the calculations in a fully spectral model, enabling the model to handle different illumination spectrums.

\noindent \textbf{Real-time Multiple Scattering} \quad The atmospheric model of Bruneton~\etal~\cite{BrunetonNeyret:2008}, which our work extends, improves the model of Schafhitzel~\etal~\cite{Schafhitzel:2007}~adding a 4\textsuperscript{th} parameter (the cosine of the angle between the Sun position and the view direction) for the lookup table and an incremental algorithm for precomputing the radiance contribution due to multiple scattering light effects. Because of the precomputation nature and the lookup tables, Bruneton's model is capable of running in real-time and produces very good results. The atmospheric model proposed does not account for molecular anisotropic, Mie scattering wavelength dependency, and the absorption of energy by molecules (\textit{e.g.}, the oxygen and ozone molecules).

In a similar model to Bruneton's model, Elek~\etal~\cite{Elek:2009} presented a real-time algorithm for the rendering of planetary atmospheres. Their model can simulate multiple scattering light effects and also uses an incremental algorithm for the precomputation of the final radiance value. Unlike Bruneton, their calculations do not take the angle between the Sun and the view direction into account, so they cannot reproduce the Earth's shadow inside the atmosphere.

Extending their previous work, Elek~\etal~\cite{Elek:2010} included the effect of scattering in water, through the use of a different attenuation coefficient, and variable atmospheric density, rewriting the Rayleigh and Mie scattering coefficients as functions of the molecular number density, utilizing fully spectral data. Similar to Bruneton and Neyret~\cite{BrunetonNeyret:2008}, Elek~\etal~do not consider Mie wavelength dependency or absorption of radiant energy.

The simulation of a correct physically-based atmospheric model is a complex task, as many physical effects impact each photon passing through the participating media. The formulation of an entirely accurate real-time model for realistic rendering of atmospheres is infeasible. However, our goal is to focus on the primary process needed to visualize the planetary atmospheres with interactive controls correctly.

\subsection{Visualization of Atmospheric Models}
The visualization of atmospheric models can be grouped into two categories that each has its own distinct requirements.

The first category concerns the visualization of observed, measured, and simulated data interacting with an atmospheric model. For instance, weather visualization and prediction consist of some of the most well-known systems and toolkits: Vis5D, VisAD, D3D, NASA's Atmosphere, the Globe program, and NOAA's Weather and Climate Toolkit~\cite{NOAAWeatherClimate}. Because these systems are aimed at the visualization of layered and volumetric data, they often lack important visual information that is crucial when visually analyzing atmospheres such as: Rayleigh and Mie scattering effects, aerial perspective, and others which help the consumer of the visualization with its context.

The software packages used for the visualization of astronomical/astrophysical data and topological information are in the second category. Among them, capable of displaying planetary surfaces and atmospheres are Uniview~\cite{klashed10uniview}, Digistar from Evans and Sutherland, Celestia, Stellarium, Gaia Sky~\cite{GaiaSky2018}, and Space Engine. These systems have different capabilities when displaying atmospheric effects. However, they lack advanced parameter control and higher-order effects such as multiple scattering, or the description of the atmospheric model in use is not available. The same category also includes the Google Maps/Earth systems~\cite{GoogleEarth:2017}, which are capable of rendering the atmospheres of Earth, Mars, and Venus in real-time, but lack controls for the user or any indication of the atmospheric model used to produce the results. Among 3D natural scene simulators, Proland is a real-time system capable of rendering Earth's atmosphere using the atmospheric model and rendering algorithm created by Bruneton~\etal~\cite{BrunetonNeyret:2008}.

\section{Background}\label{sec:background}

This section explains the fundamental concepts of radiative transfer physics and their applications in atmospheric physics rendering.

The light interactions inside a medium can be described by two distinct approaches: phenomenological~\cite{Chandrasekhar:1960, Preisendorfer:1965} and microphysical~\cite{Mishchenko:article:2006}. In the following description, we use the former approach. When the radiant energy (photons in a light ray of wavelength $\lambda$) travels through a medium, it can be absorbed, scattered, or emitted. The energy is extinct (absorbed + scattered) when photons hit a molecule or particle, or when part of the initial energy is absorbed. The fraction of absorbed energy is represented by the absorption coefficient $\beta^{abs}(\lambda)$, and described in differential form as the rate of change of energy (the radiance: the radiant flux per unit solid angle per unit projected area) per unit of length in the direction of solid angle $\vec{\omega}$ (traveling direction) at a point $\vec{x}$. The fraction of energy scattered in different directions other than $\vec{\omega}$ (also known as \textit{out-scattered} radiant energy) is described by the scattering coefficient $\beta^{scat}(\lambda)$ and, similarly to the absorption, can be formulated in differential form too:

\vspace*{-1.5mm}
\begin{equation}\label{eq:extinction}
\small
(\vec{\omega}\bullet \nabla)L(\vec{x},\vec{\omega}, \lambda) = 
	\underbrace{-(\beta^{abs}(\lambda) + \beta^{scat}(\lambda))}_{\text{extinction coefficient }\beta^{ext}(\vec{y})} 
	\,\, \cdot \,\, \underbrace{L(\vec{x},\vec{\omega}, \lambda)}_{\text{incoming radiance}}
\end{equation}

Previously scattered energy may be in-scattered into the current traveling direction, increasing the radiant energy of that ray.  The change in radiance due to in-scattering can be written in differential form by adding all radiant energy arriving at point $\vec{x}$ from the whole sphere of directions weighted by the scattering coefficient:

\begin{equation}\label{eq:in-scattering}
\small
(\vec{\omega}\bullet \nabla)L(\vec{x},\vec{\omega}, \lambda) = \int_{4\pi}{ \beta^{scat}(\lambda)P(\vec{x}, \vec{\omega}', \vec{\omega})L(\vec{x},\vec{\omega}', \lambda)d\vec{\omega}'}
\end{equation}

The term $P(\vec{x}, \vec{\omega}', \vec{\omega}) = P(cos\theta)$ is known as the phase function, which describes the angular distribution of light intensity being scattered. For isotropic scattering, radiant energy is scattered uniformly in all directions and the phase function is constant. In this paper, we consider molecules and particles that exhibit dominant scattering directions \textit{i.e.}, anisotropic scattering. Combining \autoref{eq:extinction} and~\autoref{eq:in-scattering} and adding an emission term $W_{e}(\vec{x},\vec{\omega}, \lambda)$, yields the radiative transfer equation (RTE):

\vspace*{-4.5mm}
{
  \small
\begin{multline}\label{eq:RTE}
(\vec{\omega}\bullet \nabla)L(\vec{x},\vec{\omega}, \lambda) = - (\beta^{abs}(\lambda) + \beta^{scat}(\lambda))L(\vec{x},\vec{\omega}, \lambda) + W_{e}(\vec{x},\vec{\omega}, \lambda)\\
 + \int_{4\pi}{ \beta^{scat}(\lambda)P(\vec{x}, \vec{\omega}', \vec{\omega})L(\vec{x},\vec{\omega}', \lambda)d\vec{\omega}'}
\end{multline}
}
\vspace*{-4mm}

This work does not consider media capable of emitting radiant energy, so $W_{e}(\vec{x},\vec{\omega}, \lambda) = 0$. With the rendering equation~\cite{Kajiya:1986} as a boundary condition, \autoref{eq:RTE} can be written as a purely integral equation for the radiance in the presence of participating media:

\vspace*{-6mm}
{
  \small
  \begin{multline}\label{eq:VRE}
    L(\vec{x},\vec{\omega}, \lambda) = \overbrace{T(\vec{x}, \vec{x_0})L(\vec{x_0},-\vec{\omega}, \lambda)}^{\text{extinction}}\\
    + \underbrace{\int_{\vec{x}}^{\vec{x_0}}{T(\vec{x}, \vec{y})\int_{4\pi}{ \beta^{scat}(\lambda)P(\vec{x}, \vec{\omega}', \vec{\omega})L(\vec{x},\vec{\omega}', \lambda)d\vec{\omega}'}dy}}_{\text{in-scattering}}
  \end{multline}
}
\vspace*{-2.5mm}

The \textit{transmittance} $T(\vec{x}, \vec{x_0})$ is the result of solving the differential equation~\ref{eq:extinction} and is given by:

\vspace*{-2.5mm}
\begin{equation}\label{eq:transmittance}
\small
T(\vec{x}, \vec{x_0}) = exp\Bigg(-\underbrace{\int_{\vec{x}}^{\vec{x_0}}{\beta^{ext}(\vec{y})dy}}_{\text{optical depth}}\Bigg)
\end{equation} 
\vspace*{-1.5mm}

Finally, it should be noted that the radiant energy lost by the out-scattering effect is not converted to other energy forms; this radiant energy travels further through the atmospheric medium and may be scattered into another beam of light. This way, the out-scattering light is accounted for by the extinction process and the in-scattering light.

\subsection{Atmospheric Physics}\label{sec:atmPhysics}

In the case of planetary atmospheres as the medium of interaction with the light, \autoref{eq:VRE} must the extended to take into account the contributions of the radiant energy arriving at an observer from the reflection at a point on the planet's surface. This new contributing radiance, $R(\vec{x}, \vec{\omega}, \lambda)$, is the attenuated (by \autoref{eq:transmittance}) sum of all radiant energy arriving at point $\vec{x_0}$ (hemisphere) reflected to the observer direction (solid angle $\vec{\omega}$), weighted by the \textit{bidirectional reflectance distribution function} (BRDF) $f(\vec{x}, \vec{\omega}', \vec{\omega})$ of the planet's surface and the cosine between the normal at the point $\vec{x_0}$ and the ray incoming direction $\vec{\omega}'$:

\begin{equation}\label{eq:reflected_light}
\small
R(\vec{x}, \vec{\omega}, \lambda) = T(\vec{x}, \vec{x_0}) \int_{2\pi}{ f(\vec{x}, \vec{\omega}', \vec{\omega})(\vec{n}(\vec{x_0}) \bullet  \vec{\omega}') L(\vec{x},\vec{\omega}', \lambda)d\vec{\omega}'}
\end{equation}

Small particles (molecules and atoms) and large particles (radius between $10\,\text{nm}$ and $50\,\mu \text{m}$~\cite{Thomas:2017}) are the primary sources of atmospheric absorption and scattering (we don't consider inelastic scattering (Raman scattering) in this work, \textit{i.e.}, scattered photons have the same frequency and wavelength as the incident photons). The particles' absorption and scattering coefficients in an atmosphere are described by the Rayleigh and Mie scattering theories. Rayleigh scattering~\cite{Rayleigh:1871} $\beta_R^{scat}(\lambda)$ and $P_R(\vec{x}, \vec{\omega}', \vec{\omega})$, describes the scattering of radiant energy by molecules whose radius $r$ is very small ($x \ll 1$, $\ x=(2\pi r)/\lambda$) compared to the wavelength $\lambda$ of the incident light. Mie's theory describes the scattering ($\beta_M^{scat}(\lambda)$ and $P_M(\vec{x}, \vec{\omega}', \vec{\omega})$) and absorption $\beta_M^{abs}(\lambda)$ of radiant energy by large particles~\cite{Hulst:1981, Mishchenko:2006} (the term large is related to the size of the particle radius to the wavelength of the incident light on the particle). Mie's theory is more general than Rayleigh's theory and explains the interactions between radiant energy and particles through the perspective of Maxwell's equations~\cite{Mishchenko:2002}. Combining this information yields the extended volume rendering equation for the light interacting with the atmospheric 
medium of a planet: 

\vspace*{-5mm}
{
\small
\begin{multline}\label{eq:ATM_VRE}
L(\vec{x},\vec{\omega}, \lambda) = \overbrace{ T(\vec{x}, \vec{x_0})L(\vec{x_0},-\vec{\omega}, \lambda) }^{L_0(\vec{x}, \vec{\omega}, \lambda)}\\
+ T(\vec{x}, \vec{x_0}) \cdot \overbrace{f(\vec{x}, \vec{\omega}', \vec{\omega}) \int_{2\pi}{L(\vec{x},\vec{\omega}', \lambda) \cdot (\vec{n}(\vec{x_0}) \bullet  \vec{\omega}') d\vec{\omega}'}}^{I(\vec{x}, \vec{\omega}) = f(\vec{x}, \vec{\omega}', \vec{\omega}) \cdot E(\vec{x}, \vec{\omega}, \lambda)}\\
+ \underbrace{\int_{\vec{x}}^{\vec{x_0}}{T(\vec{x}, \vec{y}) \overbrace{\int_{4\pi}{ \sum_{i \in \{R,M\}} \beta_{i}^{scat}(\lambda)P_i(\vec{y}, \vec{\omega}', \vec{\omega})L(\vec{y},\vec{\omega}', \lambda)d\vec{\omega}'}}d\vec{y}}^{J(\vec{x}, \vec{\omega})} }_{S(\vec{x}, \vec{\omega}, \lambda)}
\end{multline} 
}
\vspace*{-3mm}

\autoref{eq:ATM_VRE} represents the radiance arriving at an observer at point $\vec{x}$, looking at point $\vec{x_0}$ (direction $\vec{v}$), with the sun at position $\vec{s}$; and can be written as a series of linear operators:

\vspace*{-3.5mm}
{
  \small
\begin{align}
L(\vec{x},\vec{\omega}, \lambda) &= L_0(\vec{x}, \vec{\omega}, \lambda) + R(\vec{x}, \vec{\omega}, \lambda) + S(\vec{x}, \vec{\omega}, \lambda)\nonumber\\ 
L(\vec{x},\vec{\omega}, \lambda) &= (L_0 + R[L] + S[L])(\vec{x},\vec{\omega}, \lambda)\nonumber\\
L &= L_0 + (R + S)[L_0] + (R + S)[(R + S)[L_0]] + \ldots \nonumber\\
L &= L_0 + L_1 + L_2 + \ldots \label{eq:ATM_VRE_lin_op}
\end{align}
}
\vspace*{-5mm}

The $i^\textnormal{th}$ term in \autoref{eq:ATM_VRE_lin_op} corresponds to the light reflected and scattered $i$ times, if $i > 1$ these are known as multiple scattering terms.
 
In the following sections, we describe the atmospheric model we derived and implemented to realistically describe the Rayleigh and Mie scattering processes in atmospheres of planets and exoplanets.

\section{OpenSpace Atmospheric Model}\label{sec:method}

An atmospheric model can be described in two parts: the way the radiant energy is scattered and the way it is absorbed inside the atmosphere.

In our model for realistic visualization of atmospheres, we take a starting point in the previous works in the literature and extend it to account for the absorption of energy by different molecules, describing how the molecules of ozone and oxygen absorb radiant energy in Earth's atmosphere. We then propose a general method for the absorption of radiant energy by molecules based on results from Maxwell's equation and the adoption of complex indices of refraction~\cite{Bohren:1983}. In the case of large particles, we propose using the anomalous diffraction approximation for absorbing spheres offered by van~de~Hust~\cite{Hulst:1981} to model the extinction of radiant energy. The resulting parameterized atmospheric model includes: 

\vspace{-1mm}
\begin{itemize}
	\item representation of complex refractive indices for molecules and particles,
	\vspace{-2.5mm}
	\item general treatment of the light absorption by molecules and particles,
	\vspace{-2.5mm}	\item the correct transmittance computation for the curved path lengths of light rays inside the atmosphere (in Earth's case),
	\vspace{-2.5mm}	\item a better approximation for Rayleigh's phase function for Earth's molecular scattering and accounting for the contributions of polarizability anisotropy~\cite{Rayleigh:1871},
	\vspace{-2.5mm}	\item effective use of the Mie scattering\cite{Mishchenko:2006}  dependency of the light's wavelength to take into account the energy absorption/scattering in heavily dusty atmospheres,
	\vspace{-2.5mm}	\item the use of a wavelength-dependent Mie phase function with  
    forward and backward scattering contributions control,
	\vspace{-2.5mm}	\item use of different particle concentrations (hydrostatic equilibrium or user-defined).
	\vspace{-1mm}
\end{itemize}

We modeled the scattering and absorption of radiant energy for large particles ($10\text{nm} < r < 50\mu \text{m}$) using an extended Double Henyey-Greenstein three-parameter phase function~\cite{Kattawar:1975} to account for wavelength dependency, and a scattering coefficient dependent of the inverse square of the wavelength of the incident light~\cite{Hulst:1981}.

For Earth, we use the air mass coefficient and the distance between the ground and the top of the atmosphere to calculate the approximate length traveled by a light ray inside the atmosphere~\cite{Pickering:2002}.

In our work, we generate only clear sky images but it is possible~\cite{Schneider:2015} to account for other weather variations by adding clouds using height and type information (the clouds in our generated images are textures with no optical properties), and considering their light interaction in the atmospheric model. Other weather variations can be obtained considering the contributions of different aerosols throughout the atmospheric model. 
To simplify the proposed model and yield real-time frame rates, the diffuse radiation field contribution from the ocean is not considered.

The following sections describe these features in detail, and how to put them together to yield our proposed atmospheric model.  

\subsection{Scattering by Molecules and Atoms}

As discussed in \autoref{sec:atmPhysics}, the loss of radiant energy is described as a combination of the absorption and out-scattering effects inside the atmosphere. It is well known that different particles have different scattering and absorption properties for the different parts of the light spectrum~\cite{Chandrasekhar:1960, Hulst:1981, Bohren:1983, Thomas:2017, Keller:2013}. In this work, we are only concerned with the scattering and absorption properties of particles in a specific interval of the visible spectrum of the light spectrum, \textit{i.e.}, light with a wavelength in the interval $440\text{nm}-680\text{nm}$. Although full spectral atmospheric model rendering is possible, we restrict our computations to the light interval above, to be able to use (and validate our model with) the different tabulated data available.\\

\noindent \textbf{Rayleigh-Cabannes Scattering} \quad The \textit{Rayleigh scattering phase function}, $P_R(\theta)$, given by \autoref{eq:phase-ray}, represents a good approximation to the angular distribution of the scattered light~\cite{Thomas:2017} by small particles ($\text{radius}_{\text{particle}} \ll \lambda_{\text{incident light}}$) inside the atmosphere:

\vspace*{-2mm}
\begin{equation} \label{eq:phase-ray}
\small
P_R(\theta) = \frac{3}{4}(1+cos^2\theta) \cdot\frac{1}{4\pi}
\end{equation}
\vspace*{-2mm}

In our atmospheric model, instead of using \autoref{eq:phase-ray}, like previous atmospheric models in the literature, we used the Rayleigh phase function approximation given by Penndorf~\cite{Penndorf:1957tables}, which takes into account the anisotropy of the atmosphere molecules and effects of polarisability (through the fitting of experimental data) and use \autoref{eq:phase-ray-penndorf}.

\vspace*{-2mm}
\begin{equation} \label{eq:phase-ray-penndorf}
\small
P_R(\theta) = 0.7629(1 + 0.932\cdot cos^2\theta) \cdot\frac{1}{4\pi}
\end{equation}
\vspace*{-2mm}

Together with the Rayleigh scattering coefficient $\beta_R^{scat}(\lambda) = \beta_R^{scat}(0, \lambda)$ in \autoref{eq:rayleigh-coeff}, we can effectively describe the Rayleigh scattering process:

\vspace*{-4mm}
{
  \small
\begin{align}
\beta_R^{scat}(h, \lambda) = \frac{8\pi^3(n(\lambda)^2-1)^2}{3N\lambda^4} \cdot f(\delta) \cdot e^{-\frac{h}{H_R}}\ \big[\text{m}^{-1}\big] \label{eq:rayleigh-coeff}, \quad f(\delta) = \frac{6 + 3\delta}{6 - 7\delta}
\end{align}
}
\vspace*{-4mm}

In \autoref{eq:rayleigh-coeff}, $\lambda$, $n(\lambda)$, and $N$ are the wavelength of the incident light, the medium's refractive index, and the molecular density at sea level, respectively. Further, in \autoref{eq:rayleigh-coeff}, the exponential term describes the atmospheric density variation with height and is formally known as \emph{hydrostatic concentration function}. The scale height $H_R$ is the atmosphere's thickness if its density were uniform and $h$ is 
the atmospheric height relative to the surface.

For maximum parametrization, the molecular density and scale height can be input as functions of the mean molecular mass $M$, and the mean mass density $\rho$, of the atmosphere's gases. Thus, $H = RT/Mg$ and $N = (N_A\cdot \rho_{m})/M$ ($T$ is the temperature, and $R$, $N_A$, and $g$ are ideal gas constant, Avogadro's number, and the acceleration due to gravity on the planet's surface). This highly parameterized way to describe the atmospheric model gives the user better control over the final results.

Because we considered the existence of anisotropic molecules in the atmosphere, we applied the correction given by the function $f(\delta)$~\cite{Liou:2002} in \autoref{eq:rayleigh-coeff}. The depolarization factor $\delta$ is slightly wavelength-dependent and varies for different molecules. In this work we consider it constant in the atmosphere.
Note that the Rayleigh phase function is dependent on the wavelength $1/\lambda^4$, \textit{i.e.}, shorter wavelengths are scattered more than long wavelengths. This is the effect responsible for the bluish color of the sky in the daytime.

Our implementation of the molecular scattering gives the user the option to provide all variables needed to evaluate \autoref{eq:rayleigh-coeff} in real-time or, supply the value of the fractional part in the equation for the desired wavelengths (easily found in tabulated values in the literature~\cite{Penndorf:1957}) and only calculate the particle concentration (the exponential part in \autoref{eq:rayleigh-coeff}) for the height within the atmosphere.\\
\vspace*{-3.5mm}

\subsection{Absorption by Molecules}

Although Rayleigh~\cite{Rayleigh:1871} did not explicitly consider small particles capable of absorbing energy, we utilize his nomenclature and consider small absorbing particles as part of Rayleigh scattering theory~\cite{Kerker:1978, Bohren:1983, Mishchenko:2006}.

Not all molecules and particles have absorption properties but, when they are present, they can be described by a complex refractive index $n(\lambda) = m(\lambda) + k(\lambda) i$ for the particle/medium. To simulate the absorption effects of small particles with known complex refraction indexes, we included in our atmospheric model a general description for the absorption coefficient $\beta_{R}^{abs}(\lambda, r)$ by small particles of radius $r$, obtained by direct expansion of Mie's equations in power series~\cite{Bohren:1983, Petty:2006}:

\vspace*{-3mm}
\begin{equation}\label{eq:molec_abs_coef}
\small
\beta_{R}^{abs}(\lambda, r) = \frac{8\pi^2 r^3 N}{\lambda}\cdot \mathit{Im}\left(\frac{n(\lambda)^2 - 1}{n(\lambda)^2 + 2}\right) \cdot e^{-\frac{h}{H_R^{\text{particle}}}} \quad \big[\text{m}^{-1}\big]
\end{equation}
\vspace*{-3mm}

\autoref{eq:molec_abs_coef} is the general description of the absorption coefficient $\beta^{abs}$ in \autoref{eq:extinction} for Rayleigh scattering if the molecule's complex refraction index and concentration per unit volume ($N$) are known.

When the complex refractive index is not known or not available for individual molecules, the \textit{absorption cross-section} $\sigma^{abs}$ (probability of absorption of a photon with particular wavelength and polarization) of the particle, can be used to describe the absorption coefficient.

In the next section, we present two different ways to account for molecule energy absorption (in the visible spectrum) in Earth's atmosphere when the absorption cross-section is known.

\subsubsection{Oxygen Molecules}\label{section:oxygen}

\begin{figure}[t]
  \centering
  \fbox{\includegraphics[width=0.99\linewidth]{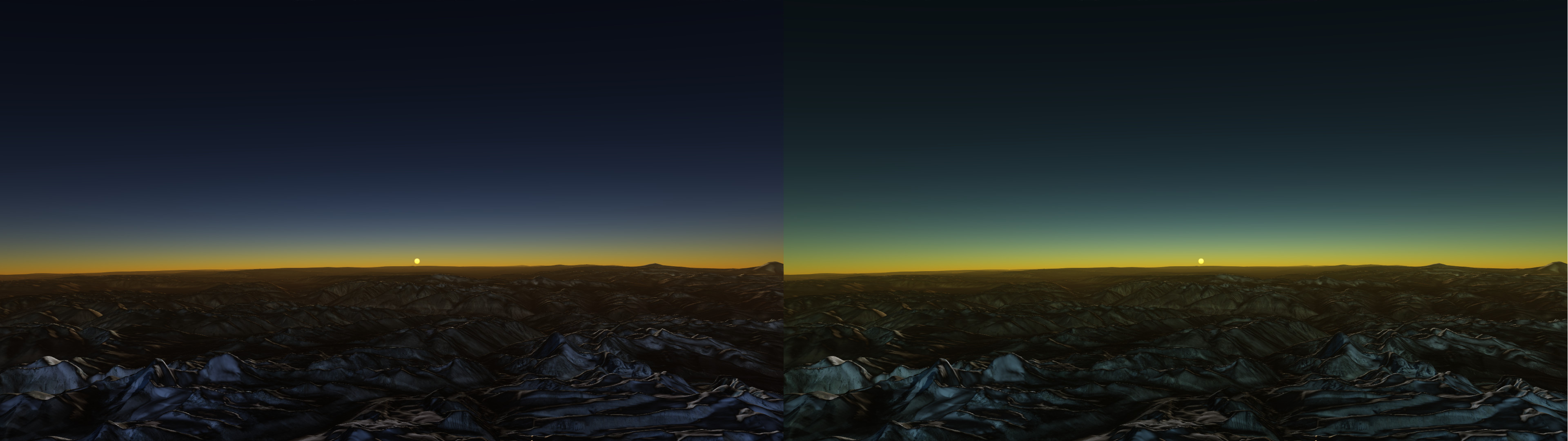}}
  \vspace*{-5mm}
  \caption{Light absorbing molecules influence the atmosphere's visual character. Our method enables the rendering of atmospheres with (left picture) or without (right picture) oxygen and ozone molecules. The ozone presence is reflected as a deeper bluish color seen above the sunsets.}
  \label{fig:ozone}
  \vspace*{-3.5mm}  
\end{figure}

Given a volume density of particles $\rho$, in a medium with absorption cross section $\sigma^{abs}$, the absorption coefficient $\beta^{abs}$ can be defined as $\beta^{abs} =\rho\sigma^{abs} $, \textit{i.e.}, the absorption cross-section area per unit volume.

It's essential to notice that the number of particles per unit of volume is not constant with the height in planetary atmospheres.  
For the oxygen molecule, in particular, we use the absorption cross-section measurements available in~\cite{Bogumil:2003}, and together with the volume density of oxygen in the air, we obtained the oxygen absorption coefficient:

\vspace*{-2mm}
\begin{equation} \label{eq:beta-oxygen}   
  \small
  \beta^{abs}_{O_2}(h, \lambda) = \sigma^{abs}_{O_2}(\lambda) \cdot \rho_{O_2}(h)\ \big[\text{m}^{-1}\big]
\end{equation}

The volume density (concentration) of oxygen particles in the dry air $\rho_{O_2}(h)$ gives the number of oxygen molecules per unit of volume at a given height (in $km$).

In Earth's atmosphere, the concentration of oxygen molecules is described by a hydrostatic concentration function multiplied by the oxygen percentage in dry air:

\vspace*{-2.5mm}
\begin{equation} \label{eq:oxygen-density}
  \small
  \rho^{\text{dry air}}_{O_2}(h) = O_2^{\%}\cdot N_0 \cdot e^{\frac{h}{H_{O_2}}}\ \big[\text{m}^{-3}\big]
\end{equation}

Once available, the $\beta^{abs}_{O_2}$ is used in the transmittance $T(\vec{x}, \vec{\omega})$ calculation to take into account the oxygen's light absorption. The results of this new absorption layer can be seen in \autoref{fig:ozone}. This process can be repeated other times with new values for the absorption cross-section area per unit of volume for different molecules (whose concentration can be described by a hydrostatic concentration function).

\vspace*{-2mm}
\subsubsection{Ozone Layer}\label{section:Ozone}

For Earth's atmosphere, the ozone molecule is particularly important. The ozone layer, the absorption layer of ozone molecules in Earth's atmosphere, absorbs light in a specific spectrum~\cite{Horshelev:2014, Serdyuchenko:2014} and thus responsible for the bluish colors during twilight~\cite{Adams:1974}.

We used previously published absorption cross-section measurements for the molecules of ozone and the ozone density of particles in Earth's atmosphere, to find the absorption coefficient for the ozone molecule as a function of height~\cite{Keller:2013, Hannelore:2013}.

The ozone's density of particles per volume is also known as the ozone concentration profile. Different from the oxygen molecules in Earth's atmosphere, it is not represented by a hydrostatic concentration function but by its specific concentration function. 

This concentration function varies during different locations and times of the year. In our atmospheric model, we use a well-known concentration profile~\cite{US_atmosphere:1976}. To obtain a continuous mathematical representation of the data for integration in \autoref{eq:transmittance}, we fitted a piecewise cubic spline $\rho_{O_3}(h)$ over the available data (see the supplementary material for details).

\vspace*{-2mm}
\begin{equation} \label{eq:beta-ozone}
  \small
  \beta^{abs}_{O_3}(h, \lambda) = \sigma^{abs}_{O_3}(\lambda) \cdot \rho_{O_3}(h)\ \big[\text{m}^{-1}\big]
\end{equation}

The results of this new absorption layer can be seen in \autoref{fig:ozone}.

\vspace*{-2mm}
\subsubsection{Absorption by Other Molecules}

Other molecules are also capable of absorbing radiant energy. For Earth, besides the $O_3$ and $O_2$ molecules, the most notable ones are $CO_2$ and $H_2O$. These molecules' absorbing cross-sections values are close to zero when the incident light's wavelength is higher than $180\text{nm}$~\cite{Shemansky:1972} and smaller than $362\text{nm}$~\cite{Lampel:2015}. Other molecules such as $H_2O_2$, $NO_2$, $NO_3$, and $CH_3$ do not occur in concentrations high enough to be taken into account in our visualizations. The molecules in Mars' atmosphere do not absorb light in the visible wavelength range~\cite{Haberle:2017}.

\subsection{Absorption and Scattering by Large Particles}

\noindent \textbf{Mie Scattering} \quad If the radius of the particle is much bigger than the wavelength of the incident photon, Rayleigh scattering no longer explains the observable values. In this case, the \textit{Mie-Debye} scattering theory must be used to describe the light scattering by those particles. 

The Mie scattering equations can be computationally expensive to compute since they are given by a direct solution of Maxwell's equations in the form of a series of Riccati-Bessel functions~\cite{Hulst:1981, Mishchenko:2006}. A much more common approach is to use approximations for the Mie scattering and absorption coefficients and the Mie phase function.

In our atmospheric model, we propose the use of a \textit{Double Henyey-Greenstein} (DHG) phase function~\cite{Kattawar:1975} with the parameters controlling the forward ($g_1$) and backward ($g_2$) scatterings and the forward-backward ratio $\alpha$ as functions of the wavelength of the incident light:

\vspace*{-4.5mm}
{
  \small
\begin{multline}
P_M(\theta, g_1(\lambda), g_2(\lambda), \alpha(\lambda)) = \bigg[ \alpha\frac{(1+g_1^2(\lambda))}{(1+g_1^2(\lambda)-2g_1(\lambda)cos\theta)^{3/2}}\\
    + (1 - \alpha)\frac{(1+g_1^2(\lambda))}{(1+g_2^2(\lambda)-2g_2(\lambda)cos\theta)^{3/2}} \bigg] \cdot\frac{1}{4\pi},
\label{eq:phase-mie}    
\end{multline}
}
\vspace*{-3.5mm}

\noindent where $g_i(\lambda) \in [-1, 1]$ and $\alpha \in [0, 1]$. The new added dependency on wavelength is direct, \textit{i.e.}, $g_i(\lambda)$ has different values for different $\lambda$ by experimental findings.
We adopted the DHG phase function because of its ability to model the scattering phase functions from Mars' dust~\cite{Chen:2019} and its high degree of parametrization.

To avoid the computational burden of calculating the Mie's equations, we use an approximation of the scattering coefficient $\beta_M^{scat}(h, \lambda)$, based on the approximations given by van de Hulst~\cite{Hulst:1981}:

\vspace*{-3mm}
\begin{equation} \label{eq:mie-scattering}
  \small
  \beta_M^{scat}(h, \lambda) = 0.434 C(T) \pi \left( \frac{2\pi}{\lambda} \right)^{\nu-2} K \cdot e^{-\frac{h}{H_M}}
\end{equation}
\vspace*{-3mm}

In \autoref{eq:mie-scattering}, the concentration factor $C(T)=(0.65T-0.65) \cdot 10^{-16}$ \cite{Morales:2017} is a function of the \textit{turbidity} $T$ (a measure of the atmosphere's haziness of a medium), and $H_M$ and $h$ are the Mie's scale height and height from the planet's surface, respectively. 
Also in \autoref{eq:mie-scattering}, $K$ is a wavelength-dependent fudge factor (an approximation of the integral of the scattering efficiency times the particles' radius as proposed by F. W. P. G\"otz in \cite{VSNGBook:1944}). The value $\nu$ is known and the \textit{Junge's exponent} and must be in the interval $[2, 6]$~\cite{Hulst:1981}. In the specialized literature and in our results we always use $3 \leq \nu \leq 4$, \textit{i.e.}, Mie's scattering has a $\lambda^{-1}$ or $\lambda^{-2}$ dependency for scattering light.\\

\vspace*{-3.5mm}
\noindent \textbf{Mie Extinction} \quad Like in calculating the Mie scattering coefficient, the Mie extinction coefficient computation can be computationally expensive. In order to avoid those heavy computations, we use the anomalous diffraction approximation for absorbing spheres proposed by van~de~Hust~\cite{Hulst:1981}.

For absorbing spheres with radius $r$ and complex refractive index $n(\lambda) = m(\lambda) + k(\lambda)i$, and incident light with wavelength $\lambda$, the Mie extinction coefficient is given by:

\vspace*{-2mm}
\begin{equation}\label{eq:Mie_extinction}
  \small
  \beta_M^{ext}(h, \lambda, r) = e^{-\frac{h}{H_R}} \int_{0}^{\infty}N(r)\pi r^2 \overline{Q_{M}^{ext}}(\lambda, \rho(r))dr ,
\end{equation} 

\noindent where the extinction efficiency factor (ratio of extinction to the geometric cross-sections) approximation $\overline{Q_{M}^{ext}}$ and the distribution of particles by radius size $N(r)$ ($C$, $a$, and $b$ are user parameters) are given by:

\vspace*{-3mm}
{
  \small
\begin{align}
\overline{Q_{M}^{ext}}(\lambda, \rho(r)) &= 2 - 4e^{-\rho(r) tan\beta}\frac{cos\beta}{\rho(r)}sin(\rho(r)-\beta) \notag\\
&- 4e^{-\rho(r)tan\beta}\left(\frac{cos\beta}{\rho(r)}\right)^2cos(\rho(r)-2\beta) \notag\\
&+ 4\left(\frac{cos\beta}{\rho(r)}\right)^2cos2\beta\label{eq:mie_ext_efficiency}\\
\rho(r) = \frac{4\pi}{\lambda}r\cdot(m(\lambda) - 1), &\quad tan\beta = \frac{k(\lambda)}{m(\lambda) - 1}, \quad
N(r) = C \cdot r^{\frac{1-3b}{b}}e^{\frac{r}{ab}} \notag
\label{eq:particle_distribution}
\end{align}
}
\vspace*{-4mm}

To keep the computational loading small, we consider only layers of particles with the same radius size, \textit{i.e.}, all particles in the unit volume have a mean radius $\overline{r}$, simplifying \autoref{eq:Mie_extinction} to:

\vspace*{-1.5mm}
\begin{equation}\label{eq:final_mie_ext}
  \small
  \beta_M^{ext}(h, \lambda, r) = \pi {\overline{r}}^2 \overline{Q_{M}^{ext}}(\lambda, \rho(r))N_{\overline{r}} \cdot e^{-\frac{h}{H_M}}
\end{equation}
\vspace*{-4.5mm}

\subsection{Non-Linear Light Path Inside the Atmosphere}\label{section:nlp}

To correctly calculate the travel distance of a light ray inside the atmosphere, the variation of medium's density with the height must be taken into account. This density variation causes atmospheric refraction, \textit{i.e.}, light does not follow a linear path through the atmosphere, but an approximately elliptical one. In our atmospheric model for \textit{Earth}, we use the Pickering's approximation \cite{Pickering:2002} to obtain the correct traveled distance for the ray inside the atmosphere. The ratio of the distance $s(\theta)$ displaced by a light ray (with zenith angle $\theta$) through the atmosphere to the perpendicular distance $r_{\text{atm}}$ between the ground and the top of the atmosphere, is defined as the \textit{air mass coefficient AM}.

In Pickering's method, the Rayleigh air mass coefficient can be determined by the apparent altitude $h_\theta = 90 - \theta$ in degrees:

\vspace*{-3mm}
{
  \small
\begin{align} \label{eq:airmass}
AM =\ \frac{s(\theta)}{r_{\text{atm}}} = \frac{1}{sin\Big((90 - \theta) + \frac{244}{(165 + 47\cdot(90 - \theta)^{1.1})}\Big)}
\end{align}
}
\vspace*{-3mm}

The length difference, when considering the correct length of a light ray inside the atmosphere $s(\theta)$, is directly reflected in the final value of the ray's transmittance. As the distance traveled increases, the extinction probability for each photon increases, too. In Earth's case, this is translated to an increase in the scattering of blue light.

\vspace*{-1mm}
\section{Implementation}

Today modern graphics hardware can interactively compute and display first order ($L_0$ and $L_1$) terms from \autoref{eq:ATM_VRE_lin_op}. To do that, the number of samples used in the integration is small, and a simplified atmospheric model (no scattering by large particles, no wavelength dependency, no complex refraction indexes, no absorption by molecules, and other simplifications) is used~\cite{Hoffman:2002, ONeil2004}. However, current GPUs are not powerful enough to interactively compute and visualize all the terms in \autoref{eq:ATM_VRE_lin_op}, and the high order terms are essential to correctly display the colors of an atmosphere during sunset and sunrise.

To visualize atmospheres in real-time with high order scattering and a complex atmospheric model, we use an incremental algorithm as presented by Elek~\cite{Elek:2009} and Bruneton~\cite{BrunetonNeyret:2008} and applied some generalizations and observations to decrease the computational load:
\begin{itemize}
	\vspace{-2.5mm}
	\item one parallel light source (star far away),
	\vspace{-2.5mm}
	\item the term $L_0$ in \autoref{eq:ATM_VRE} is the attenuated Solar radiance ($L_{Sun}$) through path $\vec{x_0}-\vec{x}$, arriving at the point $\vec{x}$, from the sun at position $\vec{s}$ (if the light source is occluded or $\vec{s} \neq \vec{v}$, then $L_0 = 0$),
	\vspace{-2.5mm}
	\item $I(\vec{x}, \vec{\omega}) = 0$ on top of the atmosphere,
	\vspace{-2.5mm}
	\item the planet's surface is considered perfectly diffuse, with constant reflectance defined by the user,
	\vspace{-2.5mm}
	\item the density of the atmosphere changes with respect to the altitude but not with respect to the latitude and longitude,
	\vspace{-2.5mm}
	\item the planet is perfectly spherical, and the atmosphere is a spherical shell symmetrical around the plane between the direction of the light source ($\vec{v}$) and the zenith vector in $\vec{x}$ (height).
	\vspace{-2.5mm}
\end{itemize}

The incremental algorithm computes one scattering order at a time ($L_i$ in \autoref{eq:ATM_VRE_lin_op}), until the $n^{\text{th}}$ iteration ($n$ is a user input), using the computation of the previous order to do it, and stores the produced data in tables to be used later during the rendering phase.

The rendering algorithm is a modified version of the algorithm by Bruneton \etal ~\cite{BrunetonNeyret:2008}. It takes into account the occlusion information for first order scattering and uses the tabulated data of the precomputation algorithm to generate the final images. The algorithm is implemented in a hybrid graphics pipeline (deferred and forward rendering) in the OpenSpace system~\cite{OpenSpace:2020, bock18openspace}, using modern C++ and OpenGL. Our implementation is available in the OpenSpace GitHub repository.

\begin{algorithm}[bth]\label{alg:rendering}
  \small
  \SetAlgoLined
  \SetKwInOut{KwIn}{Input}
  \SetKwInOut{KwOut}{Output}
  \KwIn{G-Buffer G (position, normal, and color), transmittance $T(\vec{x}, \vec{x_0})$, irradiance $E(\vec{x}, \vec{\omega}, \lambda)$, and in-scattering $S(\vec{x}, \vec{\omega}, \lambda)$ tables, atmosphere and planet radii.}
  \KwOut{Atmosphere rendered on the current attached framebuffer F.}
  
  \For{each pixel $p \in $ Image Plane}{
    Trace ray $\vec{r(t)}$ from camera position through $p$\;
    \eIf{$\vec{r(t)}$ intersects an atmosphere}{
        \eIf{observer's position $\vec{x}$ is outside the atmosphere}{
            $\vec{x} \leftarrow $ first intersection position\;
            $\vec{y} \leftarrow $ second intersection position\;
        }{
            $\vec{y} \leftarrow $ first intersection position\;
        }
        
        \eIf{distance stored in depth buffer $ < \vec{x}$ }{
            Atmosphere is occluded: $F \leftarrow $ color value from G\;
        }{
            \If{distance between $\vec{x}$ and the distance in G $ < (\vec{y}-\vec{x})$ }{
                \tcc{intersection inside the atmosphere} 
                $t \leftarrow $ updated distance traveled by the ray\;
            }
            Calculates $T(\vec{x}, \vec{r(t)})$\;
            $L_0 \leftarrow T(\vec{x}, \vec{r(t)}) \cdot SunPower$\;
            \eIf{$\vec{r(t)}$ intersects the planet's ground}{
                $R[L_0] \leftarrow T(\vec{x}, \vec{y}) \cdot c \cdot (\vec{s}\bullet\vec{n})E[L_0]$\;
            }{
                Calculate Sun's disc color\;
            }
            $S(\vec{x}\rightarrow \vec{y}, \vec{\omega}, \lambda) \leftarrow S(\vec{x}, \vec{\omega}, \lambda) - T(\vec{x}, \vec{y})S(\vec{y}, \vec{\omega}, \lambda)$\;
            
            $F \leftarrow S(\vec{x}\rightarrow \vec{y}, \vec{\omega}, \lambda) + R[L_0] + $ Sun's disc color\;
        }
    }{
        $F \leftarrow $ color value from G\;
    }
  }
  \caption{Rendering Algorithm}
 \end{algorithm}

\vspace*{-1mm}
\subsection{Precomputations}

Initially, the values for the transmittance $T(\vec{x}, \vec{x_0})$, in a perfectly spherical planet and atmosphere, are precomputed for different angles and heights and stored in a 32-bit float 2D texture. The texture is indexed by height $r$ of the observer at point $\vec{x}$ ($r = ||\vec{x}||$), and cosine $\mu$ of the angle $\theta$ between the view zenith direction and the normalized view direction $\vec{v}$, \textit{i.e.}, $\mu = cos(\theta) = (\vec{v}\bullet\vec{x})/r$. Next, the values of the irradiance $E(\vec{x}, \vec{\omega})$ and in-scattering $S(\vec{x}, \vec{\omega})$, both for all possible incoming directions $\vec{\omega}$ of directly light $L_0$, are also precomputed and stored in two separate 32-bit float texture tables (2D and 4D tables respectively). Using the initial computations for $E$ and $S$, the algorithm iterates on the number of desired scattering orders $n$ computing the increments on the irradiance and the in-scattering values.

Like the transmittance mapping, the irradiance and in-scattering tables are also mapped using the observer's position. The cosine of the angle between the view zenith and the light source position $\vec{s}$ is the second coordinate in the irradiance table, and the cosine of the angle between the light position and the view position, the second coordinate in the in-scattering table.

This mapping is directly translated to a 4D table, stored in the GPU as a 3D texture interpolated by user code and not through OpenGL's automatic hyperbolic interpolation process in the $4^{\text{th}}$ dimension.
For more details about the interpolation mapping and precomputation algorithm refer the reader to the work from Bruneton\cite{BrunetonNeyret:2008} and Yusov\cite{Yusov:2013}.

\vspace*{-1mm}
\subsection{Rendering Algorithm}

\autoref{eq:ATM_VRE} is evaluated at each pixel representing the atmosphere by applying \autoref{eq:ATM_VRE_lin_op}. During the first forward rendering pass, the position, color, and normals of each pixel are stored. In the second rendering pass the atmosphere is rendered by tracing rays from the camera position through the pixels (view direction $\vec{v}$) on the image plane and finding intersections. If the observer is outside the atmosphere, we move their position $\vec{x}$ (camera position) to the top of the atmosphere (first intersection position) and store only the second intersection. Otherwise, only the first intersection is stored.

The distance between the first intersection $\vec{y}$ and the camera position is the ray's path inside the atmosphere. If this distance is less than the distance stored in the depth buffer for the current pixel, there was an intersection inside the atmosphere and we update the distance traveled by the ray. Using this distance, we can calculate the transmittance for the ray using the data stored in the transmittance texture (storing $T(\vec{x}, \vec{x_0})$ for $\vec{x_0}$ on top of atmosphere or on the ground), applying the following identity if needed (we store the transmittance for the unoccluded ray's path): $T(\vec{x}, \vec{y}) = T(\vec{x}, \vec{x_0}) / T(\vec{y}, \vec{x_0})$. The calculated transmittance of the ray of light is multiplied by the Sun's irradiance to obtain $L_0$.

The reflection contribution, $R[L_0]$ in \autoref{eq:ATM_VRE_lin_op} is directly calculated once we know the BRDF (the constant diffuse term $c$), the normal at the intersection position (obtained from the normal buffer in the GBuffers) and the transmittance along the ray: $R[L_0] = T(\vec{x}, \vec{y}) \cdot c \cdot (\vec{s}\bullet\vec{n})E[L_0]$.

Finally, the irradiance and in-scattering tables are used to obtain the high-order terms in \autoref{eq:ATM_VRE_lin_op}. To compute $S[L_i]$, we use the fact that the scattering contribution from a ray of length $||\vec{y}-\vec{x}||$ is the same as the scattering contribution for a rays $(\vec{x_0}-\vec{x}) - (\vec{x_0}-\vec{y})$:  $S(\vec{x}\rightarrow \vec{y}, \vec{\omega}, \lambda) = S(\vec{x}, \vec{\omega}, \lambda) - T(\vec{x}, \vec{y})S(\vec{y}, \vec{\omega}, \lambda)$~\cite{ONeil:2005, Schafhitzel:2007}. This identity is reflected in the \textit{aerial perspective}: the appearance of objects as they are seen from a distance. We depicted the core steps of the rendering algorithm as a pseudo-code in Algorithm~\autoref{alg:rendering}. Once all pixel contributions have been made, we adjust the final color applying an exponential tone mapping operator. To work with various atmospheres simultaneously, we implemented a \textit{ping-pong} mechanism between two different rendering buffers into OpenSpace. Also, we use double precision on the initial intersection calculations to work with all possible huge distances in an astrovisualization system.

\vspace*{-2mm} 
\section{Results and Evaluation}

\begin{figure}[t]
  \centering 
  \fbox{\includegraphics[width=0.99\linewidth]{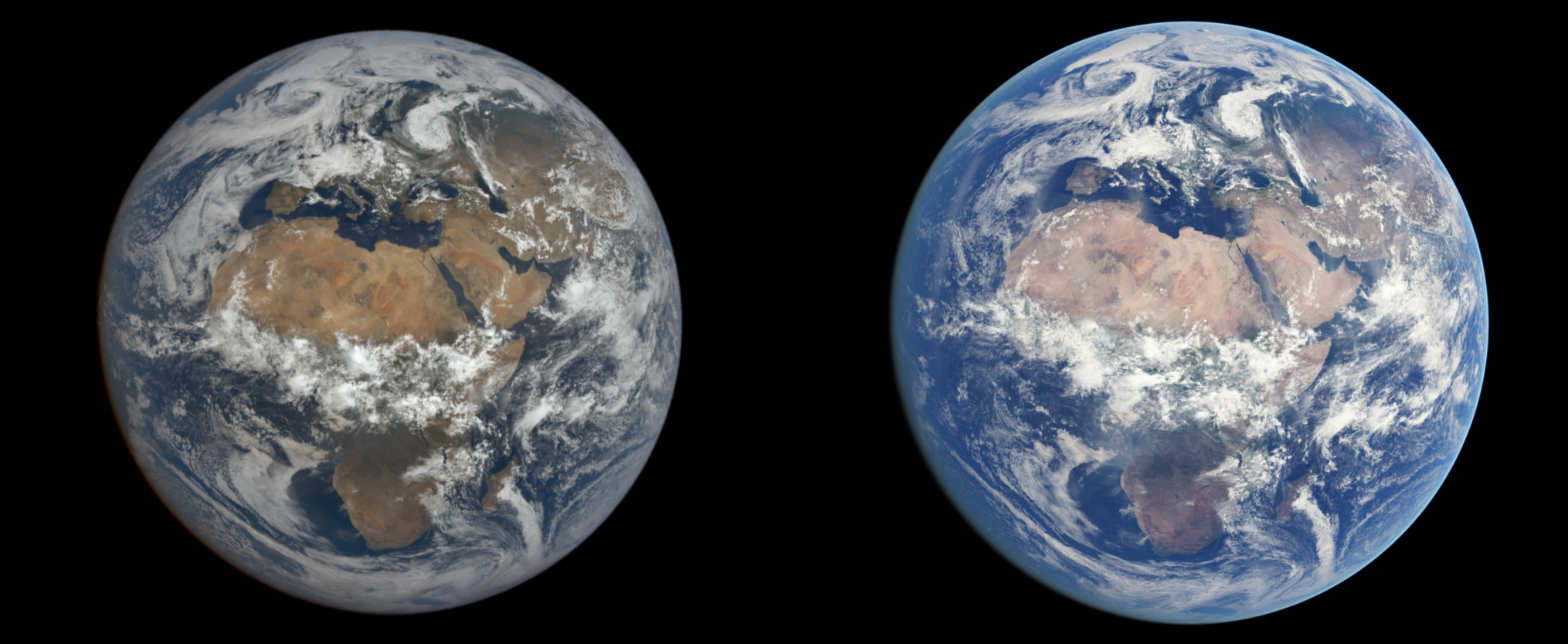}}
  \vspace*{-6mm}
  \caption{Earth's Atmosphere seen from space. (left) Taken by NASA's Earth Polychromatic Imaging Camera (EPIC), (right) generated by OpenSpace system using the our method with the parameters in \autoref{sec:Earth_ATM}.}
  \label{fig:Earth-comparison}
  \vspace*{-2mm}
\end{figure}

In this section, we describe the quality and performance of our method using experimental data from the scientific community to generate and validate the visualization of Earth's and Mars' atmospheres. 
Earth's data shows how well our atmospheric model describes a thoroughly studied and understood planet atmosphere. At the same time, Mars' data is used to generate a visualization of a planet's atmosphere is not as thoroughly understood. We also plot the generated data from our Earth's atmosphere against the CIE clear sky model~\cite{Darula:2002} for Earth's atmosphere to validate our atmospheric model. All tests were performed with an Intel Core i7-5930K and an NVIDIA GeForce GTX 1080 Ti GPU.

\vspace*{-1.5mm}
\subsection{Earth's Atmosphere}\label{sec:Earth_ATM}

Our model implementation allows the user (from domain experts to the general public) to interact with the atmospheric parameters in two modes: standard (simplified parameters) or advanced mode. 

\begin{figure}[t]
  \centering
  \fbox{\includegraphics[width=0.99\linewidth]{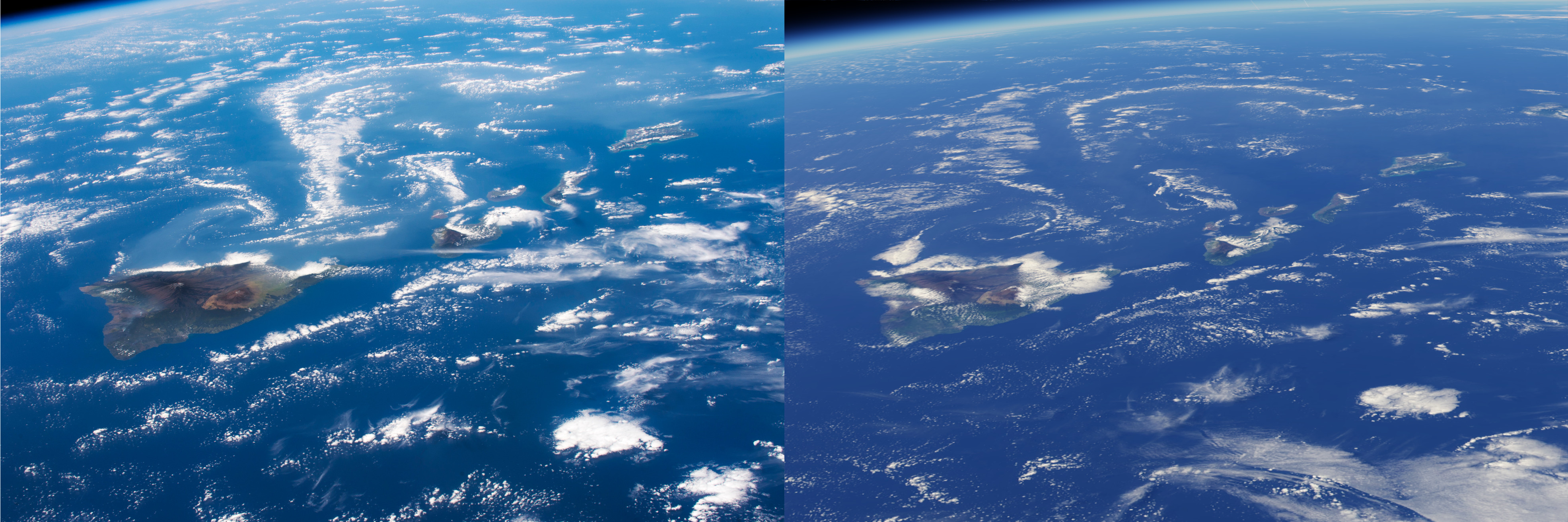}}
  \vspace*{-6mm}
  \caption{Hawaii Islands seen from the International Space Station (ISS). The picture on the left was taken from the International Space Station by NASA, while the image on the right was generated by OpenSpace using our new atmospheric model using the advanced parameters mode.}
  \label{fig:Hawaii-comp}
  \vspace*{-3.5mm}
\end{figure}

The standard parameters mode uses the Rayleigh and Mie coefficients ($\beta_R^{scat}$, $\beta_M^{scat}$ and $\beta_M^{ext}$); scale heights $H_R$ and $H_M$; and phase scattering constants as inputs in addition to common standard parameters (planetary and atmospheric radii, and the planet's diffuse term) to generate the precomputed data and final visualization of Earth's atmosphere (see \autoref{tab:parameters} for the parameters' values). 
\begin{figure*}
 \centering
 \fbox{\includegraphics[width=0.99\linewidth]{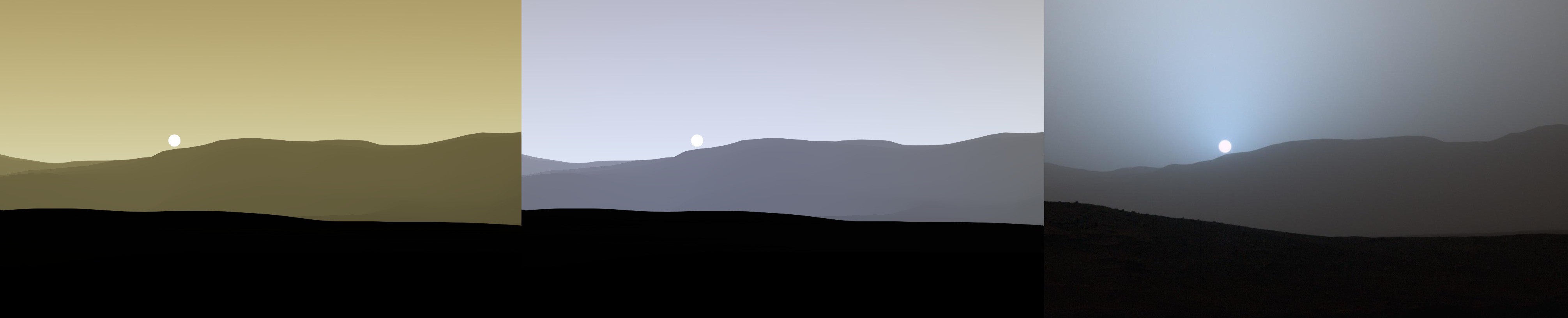}}
 \vspace*{-4mm}
 \caption{Mars sunsets: (left) Using the technique in Collienne~\cite{Collienne:2013} that simulates Mars' atmosphere colors with Rayleigh Scattering. (center) Our method that produces colors through the scattering and absorption of light by the dust particles in Mars' atmosphere. (right) Picture taken by NASA's Curiosity Mars rover (Credit NASA/JPL-Caltech/MSSS/Texas A\&M Univ.).}
 \label{fig:Mars-Rover_and_Comp_Models}
\end{figure*}
\begin{figure*} 
 \centering
  \fbox{\includegraphics[width=\textwidth]{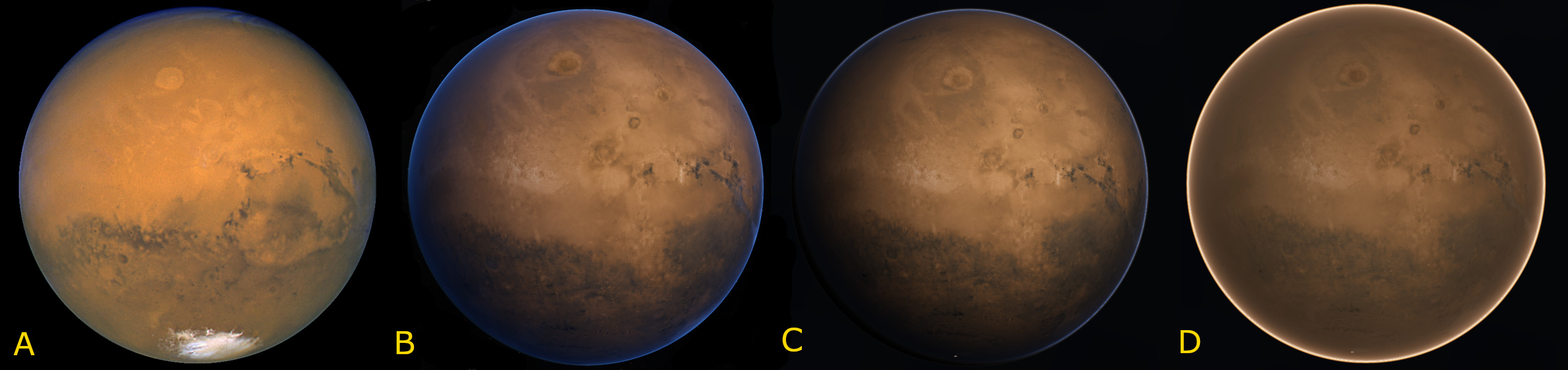}}
  \vspace*{-7mm}
  \caption{Mars viewed from space. \textbf{A:} photo taken by the Hubble telescope, \textbf{B:} visualization with Rayleigh scattering only, \textit{i.e.}, only $CO_2$ and no Mie scattering due to dust particles, \textbf{C:} atmosphere rendered with Mie effects from parameters described in \autoref{sec:Mars_ATM}, and \textbf{D:} atmosphere with same parameters as in \textbf{C} but with 20\% more particles suspended in the air. Note that the Mie scattering dominates due to the large number of particles}
 \label{fig:Mars-Ray-Mie}
 \vspace*{-2mm}
\end{figure*}
Note that in the standard mode, Mie scattering and absorption (and thus the phase functions) are not wavelength-dependent. Even in standard mode, when the number of parameters is limited, our atmospheric model can reproduce Earth's atmosphere with high accuracy (see \autoref{fig:Hawaii-comp}). 

In advanced mode, parameters like the molecular density $N$ at ground level, the complex refractive index $n(\lambda)$ for the different wavelengths, radius $r$ of the absorbing molecule for Rayleigh absorption, and all other parameters considered in \autoref{sec:method}, are used to calculate the final values of the Rayleigh and Mie coefficients (see \autoref{tab:parameters}).
The extra flexibility of the advanced mode enables the user to fine-tune most atmospheric behaviors. Our model does not allow for weather controls, seasonal and location variations of $O_3$ concentration.

\autoref{fig:teaser} shows our atmospheric model being used in exploratory scenario visualizing the effect of different concentrations of absorption particles, density of gases, and suspended number of particles. This particular approach is useful to visualize an exoplanet's atmosphere once some of its chemical composition is known.

We used the standard and advanced mode to generate pictures of the Earth and compared them to ground-truth photographs in Fig \ref{fig:Earth-comparison} and \ref{fig:Hawaii-comp}. The images are brighter than the ground-truth photographs because of the selected tone mapping operator. In clouds case, the differences in brightness are the result of no height information available at the rendering time (textured on sea level).
Also, related to the differences in the pictures, when comparing the parameter modes, the advanced mode allows for wavelength-dependency of the Mie scattering and extinction processes, and a better-fitted phase function for the Rayleigh scattering process generates more visually realistic images. 

Another challenge is that we do not have all the physical parameters (e.g., weather conditions, clouds, amount of clouds covered area, humidity, dust concentration in the air) available when the photographs were taken.
Finally, it is important to point out that our atmospheric model approximates the physical effects happening in the actual atmosphere; the Mie equations are not solved for each light interaction. 

\begingroup
\setlength{\tabcolsep}{0.2em}
\begin{table}[t]
  \centering
  \begin{tabular}{|p{1.5cm}|p{3.2cm} p{3.2cm}|}
    \hline
    \multicolumn{3}{|c|}{\textbf{Atmospheric Parameters List}} \\ [0.5ex] 
    \hline
    & \textit{Earth} & \textit{Mars}\\
    {\small $\lambda$ $[\text{nm}]$} & {\small $(680, 550, 440)$ (S, A)} & {\small $(680, 550, 440)$ (S, A)} \\
    {\small $\beta_R^{scat} [\text{m}^{-1}]$} & {\tiny $(5.1768, 12.2588, 30.5964) \cdot 10^{-6}$~\cite{Bucholtz:1995} (S)} & {\tiny $(1.2871, 3.0560, 7.6406) \cdot 10^{-3}$ (S)} \\
    {\small $\beta_M^{scat} [\text{m}^{-1}]$} & {\small $4.0 \cdot 10^{-5}$ (S)} & {\small -} \\
    {\small $\beta_M^{ext} [\text{m}^{-1}]$} & {\small $0.1 \cdot \beta_M^{scat}$ (S)} & {\small -} \\
    {\small $H_R$ [km]} & {\small 7.99575 (ICAO) (S)} & {\small 8.0~\cite{Ho:2002}}\\
    {\small $H_M$ [km]} & {\small 1.2~\cite{Morales:2017} (S)} & {\small 11.1}\\
    {\small $g_1$} & {\small 0.85 (S) (A)} & {\small (0.03, 0.4, 0.67) (A)}\\ 
    {\small $g_2$} & {\small 0.0 (S) (A)} & {\small (0.094, 0.89, 0.099) (A)~\cite{Chen:2019}}\\
    {\small $\alpha$} & {\small 1.0 (S) (A)} & {\small (0.743, 0.04, 0.01) (A)~\cite{Chen:2019}}\\
    {\tiny $N$ [part/$\text{cm}^3$] (A)} & {\small $2.68731 \cdot 10^{19}$~\cite{Penndorf:1957}} & {\small $2.8 \cdot \, 10^{29}$~\cite{Ho:2002}}\\
    {\tiny $n(\lambda)$ $[\text{nm}]$ (A)} & {\small $\approx m(\lambda)$} & {\small (1.52, 1.52, 1.52)}\\
    {\tiny $m(\lambda)$ $[\text{nm}]$ (A)} & {\tiny (1.00027598, 1.00027783, 1.00028276)~\cite{Penndorf:1957} } & {\tiny (0.001i, 0.006i, 0.013i)~\cite{Ehlers:2014}}\\
    {\small $r$ [$\mu\text{m}$] (A)} & {\small 0.0} & {\small 1.6~\cite{Ehlers:2014}}\\
    {\small turbidity $T$} & {\small $\in [2, 9]$} & {\small $\in [2, 10]$}\\
    {\small $K$} & {\small (0.0096, 0.0092, 0.0089)} & {\small (0.31, 0.16, 0.27)}\\
    {\small $\nu$} & {\small 4} & {\small 4}\\
    {\small $\delta$} & {\small 0.0279~\cite{Hosek:2012}} & {\small 0.09~\cite{Penndorf:1957}}\\
    {\small $\rho_{CO2}$} & {\small -} & {\small $2.8 \cdot \, 10^{23}$ $\text{mol}\cdot \text{m}^{-3}$~\cite{Ho:2002}}\\
    {\tiny $n_{\tiny CO2}(\lambda)$ $[\text{nm}]$ (A)} & {\small -} & {\tiny (1.00044661, 1.00045019, 1.00045558)~\cite{Ho:2002}}\\
    \hline
  \end{tabular}
  \vspace*{0.2cm}
  \caption{Atmospheric parameters used in the standard (S) and advanced mode (A) to generate Figures~\ref{fig:Earth-comparison}, \ref{fig:Hawaii-comp}, \ref{fig:Mars-Rover_and_Comp_Models}, and \ref{fig:Mars-Ray-Mie}. We used a temperature equals to $273\,\text{K}$ when possible, converting the data accordingly.}
  \label{tab:parameters}
\vspace*{-8mm}
\end{table}
\endgroup

\vspace*{-1.5mm}
\subsection{Mars' Atmosphere}\label{sec:Mars_ATM}

As in Earth, the Martian atmosphere's primary scattering process involving atoms and molecules is the Rayleigh scattering~\cite{Haberle:2017}.
Mars' atmosphere is much thinner and is composed of 95.32\% carbon dioxide~\cite{MarsFact:2018}, which is the principal scattering cross-section contributing to the Rayleigh phenomenon. Unlike Earth's atmosphere, specific values for the scattering coefficients of Mars' atmosphere are not available~\cite{Haberle:2017}. However, molecular scattering cross-sections for different wavelengths of the incident light, like $CO_2$ (obtained from experimental measurements) are available~\cite{Sneep:2005}. Together with the values of the refractive indices $n(\lambda)$~\cite{Bideau:1973} for $CO_2$, they can be used to compute $CO_2$'s Rayleigh scattering coefficients~\cite{Ityaksov:2008}. Although the authors of these works do not validate the extrapolation of values for wavelengths above $532\,\text{nm}$, the visual results agree with the theory. $CO_2$ absorbs radiant energy, but this absorption is only notable for incident light with wavelengths smaller than approximately $250\,\text{nm}$~\cite{Haberle:2017, Ityaksov:2008}, which is out of the range of wavelengths we are considering. 

We were able to calculate the absorption coefficients for the $CO_2$ in Mars' atmosphere using data from \cite{Sneep:2005, Ityaksov:2008} and the refractive index of $CO_2$. 
Given the concentration of $CO_2$ in Mars $\rho_{CO2}$, and the respective molecular scattering cross-section, we obtained the Rayleigh scattering coefficients for the $CO_2$ molecules on Mars (see \autoref{tab:parameters}). However, for an observer close to the ground inside Mars' atmosphere, the aerosol particles (dust) tend to be most prevalent [21], and Mie scattering dominates Rayleigh scattering. The dust particles in Mars (plagioclase feldspar and zeolite) contain both fine and coarse grains of hematite ($\alpha - Fe_2O_3$)~\cite{Christensen:2001}, an iron oxide with a complex refractive index ranging from the hundredths in the red spectrum to more than one in the blue spectrum. The particles with fine-grained hematite scatter more strongly longer visible wavelengths and appear red, while coarse-grained particles appear gray. This effect explains the Mars' red color and part of its atmosphere color. Hematite absorbs radiant energy too. Ehlers~\etal ~\cite{Ehlers:2014} show that the absorption properties of the hematite are responsible only for the mildly bluish appearance of the Sun's disk at sunset and not the blue glow surrounding the Sun in Mars. The blue glow is caused by the dominance of blue in near-forward scattered light inside the Martian atmosphere. 

Collienne~\etal ~\cite{Collienne:2013} presented a physically based approach based on the Rayleigh scattering process. Their strategy was to select the appropriate wavelengths for the Rayleigh scattering coefficient in a way to resemble the physical process occurring between the light and the aerosol particles within Mars' atmosphere, \textit{i.e.}; they adjusted the scattering coefficients in a way that longer wavelengths are scattered with a higher probability than shorter wavelengths. Therefore, red light with long wavelengths is scattered throughout the atmosphere, and blue light with short wavelengths is absorbed. Their approach produces images of a yellowish atmosphere with bluish sunset and sunrises. However, it lacks simulation of all atmospheric effects like the blueish limb color in Mars' atmosphere and the correct glow and blue colors for the Sun seen on Mars~(see \autoref{fig:Mars-Rover_and_Comp_Models}).

To simulate the Mie scattering and absorption properties of Mars' dust, we used the parameters in \autoref{tab:parameters}. We found some of the parameters in \autoref{tab:parameters} (like the turbidity $T$), iteratively, in order to simulate the desired weather condition (amount of suspended dust in the atmosphere) inside the Martian atmosphere.

\begin{figure}
  \centering
  \includegraphics[width=\linewidth]{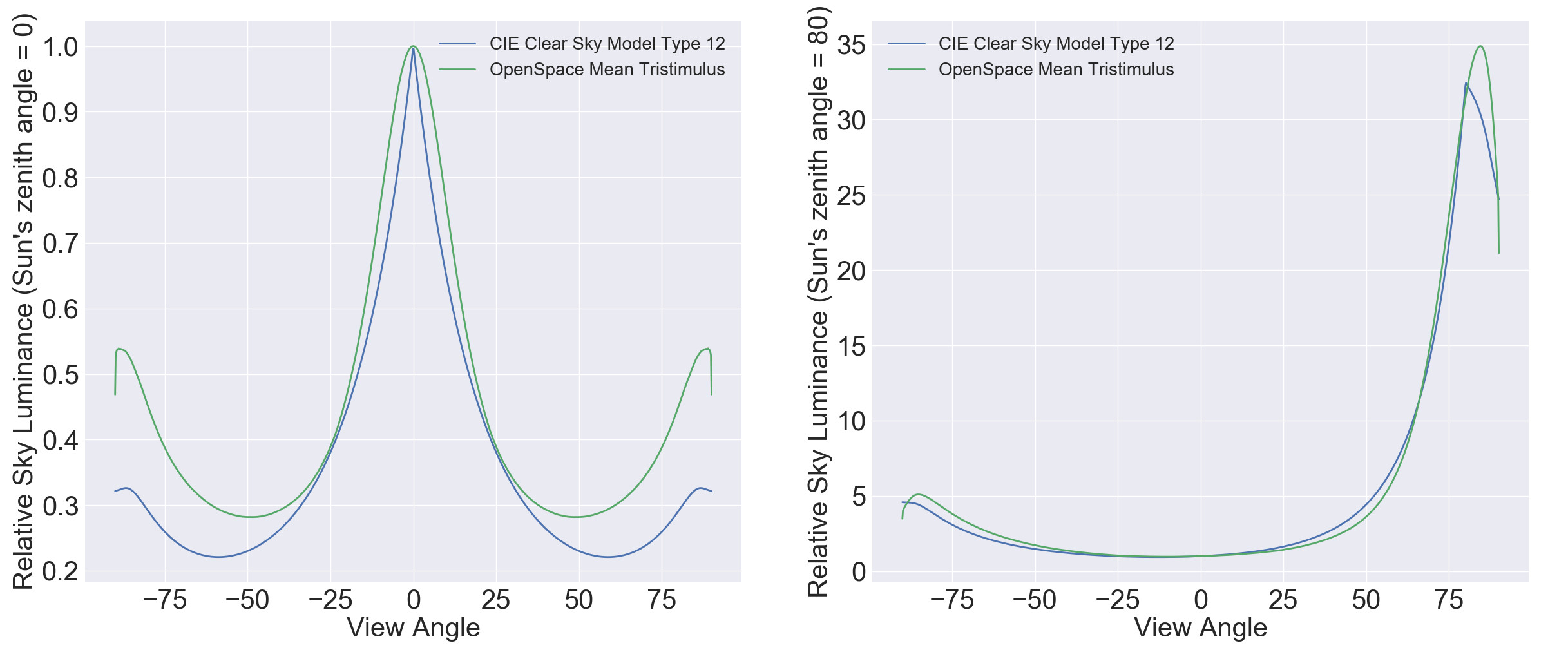}
  \vspace*{-7mm}
  \caption{The sky luminance over the zenith luminance for different Sun positions and view angles for Earth's atmosphere. The result of our advanced parametric model (green line) shows good agreement with the CIE Sky model type 12 (CIE Standard Clear Sky, low luminance turbidity) (blue line) and does not display the overestimation near the horizon displayed in other models \cite{BrunetonNeyret:2008, Preetham:1999} pointed out by Zotti \etal \cite{Zotti:2007}.}
  \label{fig:validation_curves}
  \vspace*{-6mm}
\end{figure}

To evaluate our atmospheric model and visualization of Mars, we compared the results with Mars's photographs, available at NASA's website. The middle picture in \autoref{fig:Mars-Rover_and_Comp_Models} is a visualization of the Mars' sunset obtained by our atmospheric model adapted for Mars' atmosphere. Compared to the real picture of the Martian sunset taken by the Mars Exploration Rover on the right, we can see the correct colors displayed in twilight times. When comparing the picture from the Hubble telescope of Mars' atmosphere with our generated visualization, in \autoref{fig:Mars-Ray-Mie}, we can see even the bluish limb characteristic from Mars.

In both pictures generated by our atmospheric model, we can see some differences in color and brightness when comparing to the photographs. The difference in brightness can be explained by the tone mapping parameters.  The color differences are likely due to the approximation of the Mie's physical process by the anomalous diffraction approximation and the DHG phase function.

\subsection{Comparison of Earth's Atmosphere with CIE Atmospheric Model}\label{sec:cie_model}

\autoref{fig:validation_curves} shows the results of our model plotted against the CIE clear sky model~\cite{Darula:2002} (model 12, fitted from experimental data). When compared with models from Bruneton, Preetham, and Zotti~\cite{BrunetonNeyret:2008, Preetham:1999, Zotti:2007}, our model displays the correct (approximate) behavior for higher angles. 
Our tests indicate that our atmospheric model has a better fitting on high angles because of the increased length traveled by the light, resulting from the Pickering's approximation, the Rayleigh and Mie absorption approach we proposed, and the scientific-based selection of parameters. Comparisons with varying Sun positions and view angles are available in the supplementary material.

\vspace*{-1mm}
\subsection{Performance and Memory Consumption}

\begin{figure}[b]
 \centering
 \includegraphics[width=\columnwidth]{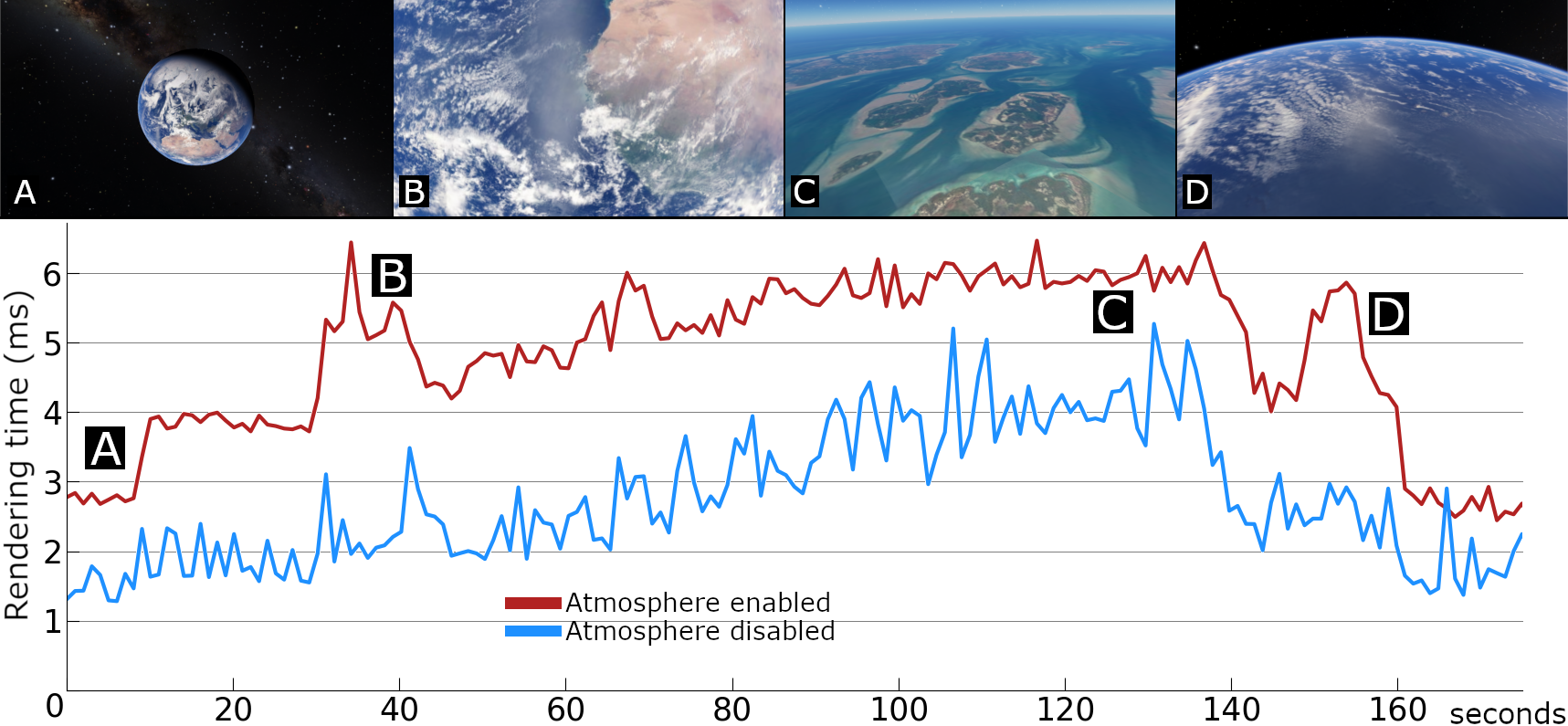}
 \vspace*{-7mm}
 \caption{The rendering times for frames of Earth during a playback of recorded flight path. The red curve shows the average frame time (in ms) using our method, while the blue curve is the time to render the same frame without an atmosphere enabled.Images A-D are stills captured at the corresponding time locations in the graph}
 \label{fig:Performance}
 \vspace*{-2mm}
\end{figure}
\begingroup
\setlength{\tabcolsep}{0.2em}
\begin{table}[t]
  \centering
  \begin{tabular}{|p{1.4cm}|p{0.8cm} p{0.8cm} p{0.8cm} p{1.4cm} p{1.4cm} p{1.4cm}|}
    \hline
    \small{\textit{Resolution}} & \small{\textit{1 ATM}} & \small{\textit{2 ATM}} & \small{\textit{3 ATM}} & \small{\textit{1 planet only}}  & \small{\textit{2 planets only}}  & \small{\textit{3 planets only}}\\
    {\small $1280\times720$} & {\small 1.81ms} & {\small 2.66ms} & {\small 5ms} & {\small 1.64ms} & {\small 1.92ms} & {\small 4.17ms}\\
    {\small $1920\times1080$} & {\small 2.63ms} & {\small 4.08ms} & {\small 6.06ms} & {\small 1.81ms} & {\small 2.56ms} & {\small 3.57ms}\\
    \hline
  \end{tabular}
  \vspace*{0.1cm}
  \caption{Average performance of our method relative to the number of visible atmospheres (ATM) in OpenSpace system.}
  \label{tab:performance}
\vspace*{-7mm}
\end{table}
\endgroup

We executed two tests to analyze the performance and scalability of our method.
First, we measured the rendering performance during playback of a recorded flight path in which the Earth was rendered from different view positions. The path was rendered with the atmosphere enabled and disabled and the results are presented in \autoref{fig:Performance} as average frame time rendered at a resolution of 1920$\times$1080 pixels.
In the second test we measured the performance when rendering multiple atmospheres simultaneously and the results are presented in \autoref{tab:performance}. As shown in Algorithm \autoref{alg:rendering}, our method is an image-based approach and, therefore, the screen resolution affects the performance linearly with the number of pixels as well as linearly with the number of simultaneous atmospheres. Both \autoref{fig:Performance} and \autoref{tab:performance} report the baseline rendering without the atmosphere to highlight the impact of components that are not directly related to the atmosphere, such as the terrain rendering which causes the high-frequency noise in the performance measurements~\cite{bladin17globe}. 

The memory consumption is directly related to the size of the textures storing the precomputations and the framebuffer size (GBuffer). In a window resolution of 1920$\times$1080 pixels, the GBuffer structures consume approximately $71.2\,\text{MB}$ of VRAM ($1920\times1080 \times 3 \text{(buffers)} \times 3 \text{(RGB)} \times 32 \text{(bits)}$), while the precomputation textures consume around $12.2\,\text{MB}$ of VRAM (transmittance texture: 256$\times$64, irradiance texture: 64$\times$32, and in-scattering texture: ($256 = 32 \cdot 8$)$\times 128 \times 32$).

\vspace*{-1mm}
\section{Application} \label{sec:application}

\begin{figure}
 \centering
 \fbox{\includegraphics[width=0.99\linewidth]{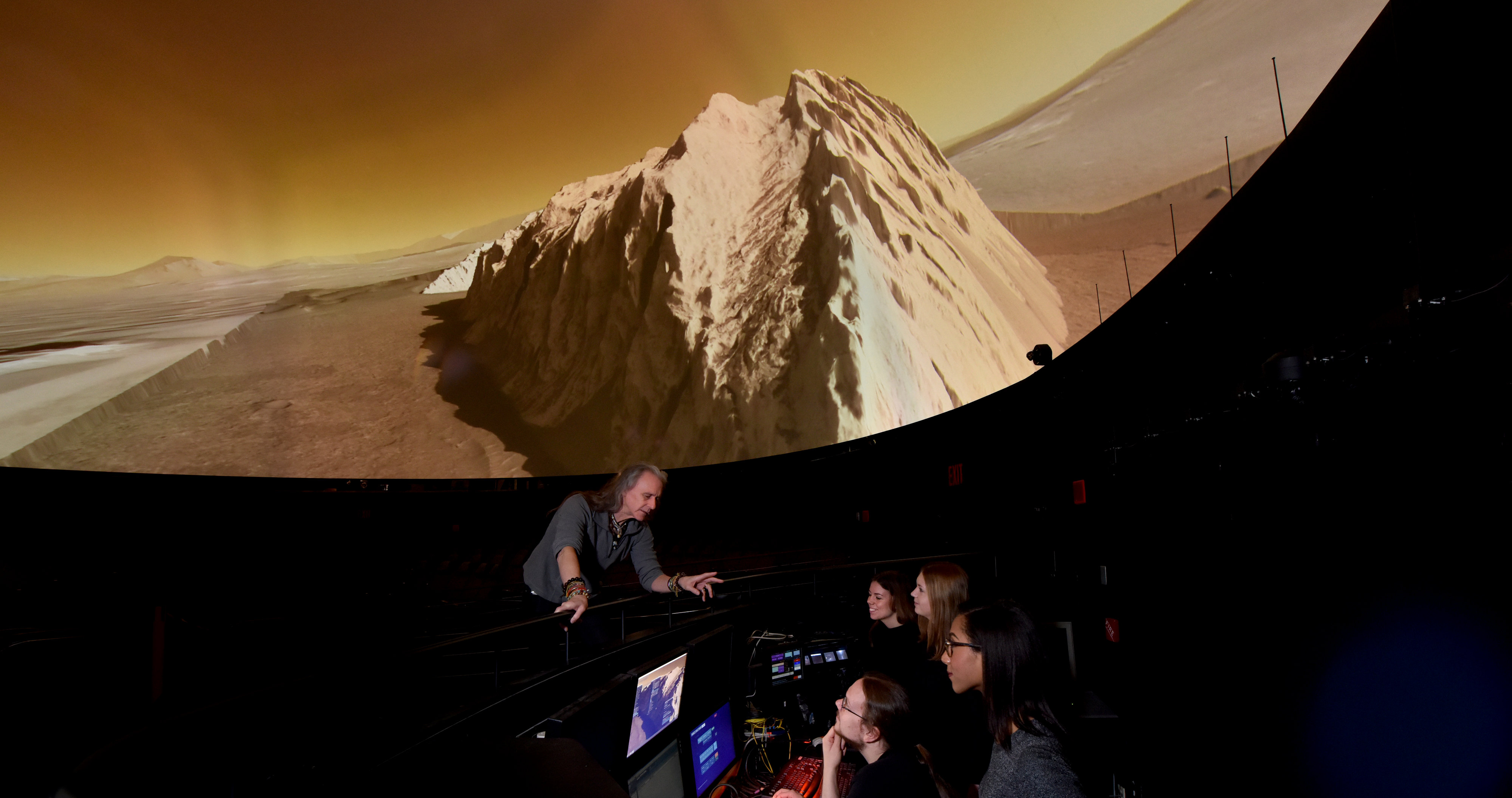}}
 \vspace*{-6mm}
 \caption{Use of OpenSpace in an interactive Science Communication situation at the Hayden Planetarium showing Martian surface features.}
 \label{fig:application-mars}
 \vspace*{-4mm} 
\end{figure}
As described in the introduction, the use of an atmospheric model serves multiple purposes and is an essential component for the use of OpenSpace as a science communication platform as well as a tool for scientific exploration. An example of a science communication situation is shown in~\autoref{fig:application-mars}, where students are exploring the martian surface in a dome theater. The atmospheric effects shown on the dome create an immersive experience and thus contribute to a high level of engagement and involvement. The high performance of the model presented in this paper supports interactive frame rates at 8K resolution, which is needed in modern multi projector planetarium configurations. Rendering of realistic looking atmospheres is thus an essential component in the wide spread use of interactive astrovisualization in science communication at planetariums and science centers. 

Another example of science exploration and communication use is dome sessions with the Mars 2020 rover science definition team chaired by Jack Mustard, Professor of Earth, Environmental, and Planetary Sciences and Professor of Environmental Studies, Brown University, who used OpenSpace for exploration and presentations of the criteria for landing spot selections. The atmosphere provided visual context when reviewing different potential landing sites.

We also foresee that the OpenSpace atmosphere model will enable rapid production of rendering of planets in wide spread science communication  of the continued exploration of the solar system conducted by science institutions and space agencies. An exciting future use is the production of realistic looking imagery of exoplanets based on scientific data collection and simulations.

\vspace*{-1.7mm}
\section{Conclusion and Future Work}

In this work we presented a new method for physically-based visualization of planetary atmospheres. Our method uses a new advanced atmospheric model capable of simulating the absorption of radiant energy by small and large particles comparable to the wavelength of the incident light, modeling different properties of dust particles based on their refractive index and radius. In Earth's case, our atmospheric model can simulate non-linear paths of the light inside the atmosphere.\\
Although our method requires modeling absorption by particles of the same size, more than one type of particle (size) can be added by simple addition of absorption coefficients.

We validated our method using data from CIE and visual comparisons with pictures taken from Mars' atmosphere by NASA. Finally, our method is designed to be used by domain experts, students, and the general public, generating images in real-time.

Future work includes adding the sun's limb darkening effect, adding the light contribution of the stars and Moon, improving the reflection BRDF (Hapke's BRDF~\cite{Hapke:2002}), and adding weather effects and clouds. More ambitious improvements include handling polarization effects for each light interaction and using Mie's equations to calculate the exact scattering phase function.

\vspace*{-1.7mm}
\acknowledgments{
This work was supported by NASA Cooperative Agreement Notice under grant NNX16AB93A, the Knut \& Alice  Wallenberg  Foundation  through the WISDOME  project, VR grant number 2015-05462, the Swedish e-Science Research Centre, and NSF awards: CNS-1229185, CCF-1533564, CNS-1544753, CNS-1626098, CNS-1730396, CNS-1828576. The source code is available at \texttt{https://github.com/OpenSpace/OpenSpace}.}

\newpage
\balance
\bibliographystyle{abbrv-doi}

\bibliography{atm-paper}
\end{document}